\documentclass[aps,superscriptaddress]{revtex4}
\usepackage{amssymb,amsmath,epsfig}
\usepackage{subfigure}
\usepackage[colorlinks=true, pdfstartview=FitV, linkcolor=blue, citecolor=red, urlcolor=magenta, breaklinks=true]{hyperref}
\begin{document}
\title{Noncommutative correction to the entropy of BTZ black hole with GUP}
\author{M. A. Anacleto}
\email{anacleto@df.ufcg.edu.br}
\affiliation{Departamento de F\'{\i}sica, Universidade Federal de Campina Grande
Caixa Postal 10071, 58429-900 Campina Grande, Para\'{\i}ba, Brazil}
\author{F. A. Brito}
\email{fabrito@df.ufcg.edu.br}
\affiliation{Departamento de F\'{\i}sica, Universidade Federal de Campina Grande
Caixa Postal 10071, 58429-900 Campina Grande, Para\'{\i}ba, Brazil}
\affiliation{Departamento de F\'isica, Universidade Federal da Para\'iba, Caixa Postal 5008, 58051-970 Jo\~ao Pessoa, Para\'iba, Brazil}
\author{B. R. Carvalho}
\email{brunorego93@gmail.com}
\affiliation{Departamento de F\'{\i}sica, Universidade Federal de Campina Grande
Caixa Postal 10071, 58429-900 Campina Grande, Para\'{\i}ba, Brazil}
\author{E. Passos}
\email{passos@df.ufcg.edu.br}
\affiliation{Departamento de F\'{\i}sica, Universidade Federal de Campina Grande
Caixa Postal 10071, 58429-900 Campina Grande, Para\'{\i}ba, Brazil}

\begin{abstract} 
We investigate the effect of noncommutativity and quantum corrections to the temperature and entropy of a BTZ black hole based on a Lorentzian distribution with the generalized uncertainty principle (GUP). 
To determine the Hawking radiation in the tunneling formalism we apply the Hamilton-Jacobi method by using the 
Wentzel-Kramers-Brillouin (WKB) approach.
In the present study we have obtained logarithmic corrections to entropy due to the effect of noncommutativity and GUP. We also address the issue concerning stability of the non-commutative BTZ black hole by investigating its modified specific heat capacity.

\end{abstract}
\maketitle
\pretolerance10000
\section{Introduction}
The study of three-dimensional gravity has been extensively explored in the literature~\cite{Carlip:1995qv}. 
It has become an excellent laboratory for a better understanding of the fundamentals of classical and quantum gravity and also to  explore some ideas behind the AdS/CFT correspondence~\cite{AdS}. 
This special attention in three-dimensional gravity has been mainly due to the discovery of the black hole solution in 2 + 1 dimensions~\cite{BTZ}.
In addition, generalizations of the Ba\~{n}ados-Teitelboim-Zanelli (BTZ) black hole solution have also been constructed considering coupling with a dilaton/scalar field~\cite{Chan:1995wj,Martinez:1996gn,Chan:1996rd}. 
In recent years, the implementation of noncommutativity in black hole physics has been extensively explored 
(for a review see~\cite{Nicolini:2008aj}). 
In~\cite{Nicolini:2005vd} the authors have introduced a noncommutative Schwarzschild black hole solution in four dimensions. 
As shown in~\cite{Nicolini:2005vd}, one way to incorporate noncommutativity into General Relativity is to modify the source of matter.
Thus, noncommutativity is introduced by replacing the point-like source term with a Gaussian
distribution --- or otherwise by a Lorentzian distribution~\cite{Nozari:2008rc}. 
In addition, noncommutativity in BTZ black hole has also been introduced 
in~\cite{Banados:2001xw,Chang,Liang:2012vx}. 
In~\cite{Anacleto:2014cga} the gravitational Aharonov-Bohm effect due to BTZ black hole in a noncommutative background has been analyzed.
The process of massless scalar wave scattering by a noncommutative black hole via Lorentzian smeared
mass distribution has been explored in~\cite{Anacleto:2019tdj}.
The thermodynamic properties of BTZ black holes in noncommutative spaces have been studied 
in~\cite{Rahaman:2013gw,Sadeghi:2015nzp,Anacleto:2015kca,Kawamoto:2017ufe}.  

It is well known that string theories, loop quantum gravity and noncommutative geometry presents important elements for the construction of a compatible theory of quantum gravity. 
Furthermore, these theories have a common feature, which is the appearance of a minimum length on the order of the Planck scale.
This therefore leads to a modification of the Heisenberg uncertainty principle, which is called the generalized uncertainty principle (GUP)~\cite{vag1,vag2,vag3}.
In recent years, several works have been devoted to investigating the effect of GUP on computing Hawking radiation from black holes in 2 + 1 dimensions. 
In this sense, the Hamilton-Jacobi method via the WKB approach to calculate the imaginary part of the action is an effective way of investigating Hawking radiation as a process of tunneling particles from a black hole~\cite{Anacleto:2019rfn,Gecim:2017vgx,Gecim:2017lxx,Anacleto:2015mma,Anacleto:2015rlz,Anacleto:2014apa,Silva:2012mt}.
In~\cite{Singh:2019oxh} the effect of GUP on Hawking radiation from the BTZ black hole has been investigated using the modified Dirac equation.
Hawking radiation has been analyzed in~\cite{Gecim:2017zid}, by considering the Martinez-Zanelli black hole in 2 + 1 dimensions~\cite{Martinez:1996gn} and using the Dirac equation modified by the GUP.
By applying the quantum tunneling formalism, Hawking radiation from a new type of black hole in 2 + 1 dimensions has also been studied in~\cite{Gecim:2017nbh}, and in~\cite{Gecim:2018sji} was explored the Hawking radiation of charged rotating BTZ black hole with GUP. Moreover, 
in~\cite{Sadeghi:2016xym,Meitei:2018mgo} the entropy of the BTZ black hole with GUP has been determined, and 
in~\cite{Maghsoodi:2020gwg}, by adopting a new principle of extended uncertainty, its effect on the thermodynamics of the black hole has been examined.

The purpose of this paper is to investigate the effect of noncommutative and quantum corrections coming from the GUP for the calculation of the temperature and entropy of a BTZ black hole based on a Lorentzian distribution,  by considering the tunneling formalism framework through the Hamilton-Jacobi method. 
Thus, Hawking radiation will be computed using the WKB approach.
Therefore, we show that the entropy of the BTZ black hole presents logarithmic corrections due to the both aforementioned effects.

The paper is organized as follows. In Sec.~\ref{BTZLD} we consider noncommutative corrections for the BTZ black hole metric implemented via Lorentzian mass distribution. 
We also have applied the Hamilton-Jacobi approach to determine noncommutative corrections for 
Hawking temperature and entropy.
In Sec.~\ref{HTEGUP} we consider the GUP to compute quantum corrections to Hawking temperature and entropy and also briefly comment on the correction of the specific heat capacity at constant volume.
In Sec.~\ref{concl} we make our final considerations.

\section{Noncommutative corrections to the BTZ black holes}\label{BTZLD}
In this section we introduce the noncomutativity by considering a Lorentzian mass distribution, 
given by~\cite{Nicolini:2005vd,Nozari:2008rc,Liang:2012vx,Anacleto:2019tdj}
\begin{eqnarray}
\rho_{\theta}(r)=\frac{M\sqrt{\theta}}{2\pi(r^2+\theta)^{3/2}},
\end{eqnarray}
where $ \theta $ is the noncommutative parameter with dimension of ${length}^2$ and $ M $ is the total mass diffused throughout the region
of linear size $ \sqrt{\theta} $. 
In this case the smeared mass distribution function becomes~\cite{Liang:2012vx} 
\begin{eqnarray}
{\cal M}_{\theta}&=&\int_0^r\rho_{\theta}(r)2\pi r dr=M\left(1-\frac{\sqrt{\theta}}{\sqrt{r^2+\theta}}\right),
\\
&=&M-\frac{M\sqrt{\theta}}{r} + {\cal O}(\theta^{3/2}). 
\end{eqnarray}

By considering the above modified mass, the metric of noncommutative BTZ black hole is given by
\begin{eqnarray}
\label{mbtzj}
ds^{2} = -{\cal F}(r)dt^{2} + {\cal F}(r)^{-1}dr^{2}+r^{2}\left(d\phi-\frac{J}{2r^{2}}dt\right) ^{2},
\end{eqnarray}
where
\begin{eqnarray}
&&{\cal F}(r)=-{\cal M}_{\theta}+ \frac{r^{2}}{l^{2}} +\frac{J^2}{4r^2} 
= -M +\frac{M\sqrt{\theta}}{r} + \frac{r^{2}}{l^{2}} +\frac{J^2}{4r^2}.
\label{F1}
\end{eqnarray}
{Note that the metric obtained by noncommutative correction is different from the metric in~\cite{Carlip:1995qv}. 
A term, $ M\sqrt{\theta}/r $, of the Schwarzschild type is generated due to noncommutative correction. 
Our metric shows similarities with the metric obtained in~\cite{Chan:1995wj,Martinez:1996gn} and also with one of the classes of solutions found in~\cite{Chan:1996rd} with a dilaton/scalar field. }

We shall now analyze the non-rotating  case ($ J=0 $), so the metric (\ref{mbtzj}) becomes
\begin{eqnarray}
ds^{2} = -f(r)dt^{2} + f(r)^{-1}dr^{2} + r^{2}d\phi ^{2},
\label{met-BTZ}
\end{eqnarray}
where
\begin{eqnarray}
&&f(r) = -M +\frac{M\sqrt{\theta}}{r} + \frac{r^{2}}{l^{2}}.
\label{f}
\end{eqnarray}
The horizons are found by solving the equation
\begin{eqnarray}
\label{hf}
f(r)=-M +\frac{M\sqrt{\theta}}{r} + \frac{r^{2}}{l^{2}}=0,
\end{eqnarray}
which is equivalent to solving a cubic equation
\begin{eqnarray}
r^3 - Ml^2 r + M l^2\sqrt{\theta}=0.
\end{eqnarray}
The roots of this cubic equation are given by~\cite{Visser:2012wu}
\begin{eqnarray}
\label{cr}
r=2\sqrt{\frac{l^2 M}{3}}\sin\left[\frac{1}{3}\sin^{-1}\left(\frac{3}{2}
\sqrt{\frac{3\theta}{l^2 M}} \right) 
+ \epsilon \frac{2\pi}{3} \right], \quad\quad \epsilon \in\{0,\pm 1\}.
\end{eqnarray}
The three roots for $ \epsilon =1, 0, -1$, up to first order in $\sqrt{\theta}$, are given respectively by
\begin{eqnarray}
&&\tilde{r}_h=r_h-\frac{\sqrt{\theta}}{2} + \cdots,
\\
&&r_c=\sqrt{\theta} + \cdots,
\\
&&r_v=-r_h-\frac{\sqrt{\theta}}{2} + \cdots,
\end{eqnarray}
where $ r_h=\sqrt{l^2 M} $, $ \tilde{r}_h $ is the event horizon, $ r_c $ the cosmological horizon and $ r_v $ the virtual (unphysical) horizon.
From Eq. (\ref{hf}) we obtain the mass of the noncommutative black hole, up to first order in $\sqrt{\theta}$, that is given by
\begin{eqnarray}
\label{mrh}
M&=&\frac{\tilde{r}_h^{2}}{l^{2}} +\frac{\tilde{r}_h\sqrt{\theta}}{l^2} + \cdots.
\end{eqnarray}

In order to compute the Hawking temperature we use the Klein-Gordon equation for a scalar field $\Phi$  in the curved space given by
\begin{eqnarray}
	\left[\frac{1}{\sqrt{-g}}\partial _{\mu}(\sqrt{-g}g^{\mu \nu}\partial _{\nu}) - \frac{m^{2}}{\hbar ^{2}}\right] \Phi = 0 ,
	\label{Klein Gordon}
\end{eqnarray}
where $ m $ is the mass of a scalar particle. In the sequel we apply the WKB approximation
	\begin{eqnarray}
	\Phi = \exp\left[\frac{i}{\hbar}I(t,r,x^{i})\right],
	\label{WKB}
	\end{eqnarray}
such that we obtain	
	\begin{eqnarray}
	g^{\mu\nu}\partial _{\mu}I\partial _{\nu}I + m^{2} = 0.
	\label{KG-WKB}
	\end{eqnarray}
By applying the metric (\ref{met-BTZ}) in the above equation we have
	\begin{eqnarray}
	-\frac{1}{f(r)}(\partial _{t}I)^{2} + f(r)(\partial _{r}I)^{2} + \frac{1}{r^{2}}(\partial _{\phi}I)^{2} + m^{2} = 0.
	\label{KG-WKB2}
	\end{eqnarray}
Now	we can write the solution of equation (\ref{KG-WKB2}) as follows
	\begin{eqnarray}
	I = -Et + W(r) + J_{\phi}\phi,
	\label{I}
	\end{eqnarray}
where 
\begin{eqnarray}
\partial_t I=-E, \quad \partial_r I=\frac{dW(r)}{dr}, \quad \partial_{\phi}I=J_{\phi},
\end{eqnarray}	
being $J_{\phi}$ a constant. 
By substituting (\ref{I}) into equation (\ref{KG-WKB2})
and solving for $W(r)$ the classical action is written as follows:
\begin{eqnarray}
I = -Et + \int dr \frac{\sqrt{E^2 - f(r)\left(\frac{J_{\phi}^2}{r^2} + m^2\right)}}{f(r)}
 + J_{\phi}\phi.
\label{I2}
\end{eqnarray}
Next, in the regime near the event horizon of the noncommutative BTZ black hole, $ r\rightarrow \tilde{r}_{h} $, we can write $ f(r)\approx\kappa (r - \tilde{r}_{h}) $ and so the spatial part of the action function reads
\begin{eqnarray}
	W(r) =\frac{1}{\kappa}\int dr \frac{\sqrt{E^2 - \kappa(r - \tilde{r}_{h}) \left(\frac{J_{\phi}^2}{r^2} 
	+ m^2\right)}}{(r-\tilde{r}_{h})}
	=\frac{2\pi i}{\kappa}E,
	\label{W}
\end{eqnarray}
where $ \kappa $ is the surface gravity of the noncommutative BTZ black hole given by
	\begin{eqnarray}
	\kappa = f'(\tilde{r}_{h}) = \frac{2\tilde{r}_{h}}{l^2}-\frac{M\sqrt{\theta}}{\tilde{r}^2_h}.
	\label{k}
	\end{eqnarray}
The next step is to determine the probability of tunneling for a particle with energy $ E $ and for this we use the following expression		
	\begin{eqnarray}
	\Gamma \simeq \exp[-2{\rm\, Im}(I)]=\exp\left(-\frac{4\pi E}{\kappa}\right).
	\label{Gamma}
	\end{eqnarray}
In order to calculate the Hawking temperature of the noncommutative BTZ black hole we can compare equation (\ref{Gamma}) with the Boltzmann factor  $\exp({-{E}/{\tilde{T}_{H}}})$, so we can find
\begin{eqnarray}
\label{ThB}
	{\tilde T}_{H} = \frac{\kappa}{4\pi}
	=\frac{\tilde{r}_h}{2\pi l^2}-\frac{M\sqrt{\theta}}{4\pi\tilde{r}^2_h}.
	\label{TH}
\end{eqnarray}
Moreover, the above result can be rewritten in terms of $r_h=\sqrt{l^2M}$ as follows 
\begin{eqnarray}
\label{tht}
{\tilde T}_H=\frac{r_{h}-\sqrt{\theta}/2}{2\pi l^2}-\frac{M\sqrt{\theta}}{4\pi{r}^2_h}
=T_{h}-\frac{\sqrt{\theta}}{4\pi l^2}-\frac{M\sqrt{\theta}}{4\pi{r}^2_h}.
\end{eqnarray}	
Therefore, the result above shows that the Hawking temperature is modified due to the presence of the noncommutative parameter $\theta$.
Note that when we take $ \theta=0 $, we recover the temperature of the commutative BTZ black hole, which is $ T_h={{r}_h}/{(2\pi l^2)}$.


At this point, we are prepared to go further.  Let us now consider the noncommutative BTZ black hole in the rotating regime ($ J\neq 0 $). 
Now the line element of equation (\ref{mbtzj}) can be written in the form	
\begin{eqnarray}
\label{mdiag}
ds^{2} = -{\cal F}(r)dt^{2} + {\cal F}(r)^{-1}dr^{2}+r^{2}d\varphi ^{2},
\end{eqnarray}
where
\begin{eqnarray}
&&{\cal F}(r) = -M +\frac{M\sqrt{\theta}}{r} + \frac{r^{2}}{l^{2}} +\frac{J^2}{4r^2},
\label{Q1}
\\
&& d\varphi=d\phi- \frac{J}{2r^{2}}dt.
\label{tc}
\end{eqnarray}
Thus, to find the horizons we have to solve
\begin{eqnarray}
\label{Fr1}
{\cal F}(r)= -M +\frac{M\sqrt{\theta}}{r} + \frac{r^{2}}{l^{2}} +\frac{J^2}{4r^2}=0,
\end{eqnarray}
which is equivalent to solving a quartic equation
\begin{eqnarray}
r^4 - l^2 M r^2 + \frac{l^2 J^2}{4}+ l^2 M \sqrt{\theta} r=0.
\end{eqnarray}
We can now rewrite this equation as follows~\cite{Visser:2012wu}
\begin{eqnarray}
\label{eqh}
(r^2-r^2_{+})(r^2-r^2_{-}) + l^2 M \sqrt{\theta}r=0,
\end{eqnarray}
that for $ \theta =0 $ we have
\begin{eqnarray}
r^2_{\pm}=\frac{l^2 M}{2}\left[1\pm \sqrt{1-\left(\frac{J}{M l}\right)^2}\right],
\end{eqnarray}
where $ r_+ $ is the outer event horizon and $ r_- $ is the inner event horizon of the commutative BTZ black hole. Now rearranging the equation (\ref{eqh}) in the form
\begin{eqnarray}
r^2=r^2_{\pm} - \frac{r^2_h \sqrt{\theta} r}{r^2 - r^2_{\mp}},
\end{eqnarray}
where $ r_h=\sqrt{l^2 M} $, we can solve it perturbatively. 
So, in the first approximation we get the event horizon
\begin{eqnarray}
\tilde{r}^2_{+}\approx r^2_{+} + \frac{r^2_h \sqrt{\theta} r_{+}}{r^2_{-} - r^2_{+}},
\end{eqnarray}
or by keeping terms up to first order in $ \sqrt{\theta} $, we obtain 
\begin{eqnarray}
\tilde{r}_{+}=r_{+} + \frac{r^2_h \sqrt{\theta}}{2(r^2_{-} - r^2_{+})}+\cdots,
\end{eqnarray}
for the outer horizon. For the internal horizon we have
\begin{eqnarray}
\tilde{r}^2_{-}\approx r^2_{-} - \frac{r^2_h \sqrt{\theta} r_{-}}{r^2_{-} - r^2_{+}},
\end{eqnarray}
so that for $ \tilde{r}_{-} $, we find
\begin{eqnarray}
\tilde{r}_{-}=r_{-} - \frac{r^2_h \sqrt{\theta}}{2(r^2_{-} - r^2_{+})}+\cdots.
\end{eqnarray}

In order to determine the Hawking temperature for the case of the rotating black hole, we can follow the same steps as presented above and so for the tunneling probability we have
\begin{eqnarray}
\Gamma=\exp[-4\pi E/\bar{\kappa}], 
\end{eqnarray}
where the surface gravity is given by
\begin{eqnarray}
\bar{\kappa}={\cal F}^{\prime}(\tilde{r}_{+})=\frac{2\tilde{r}_{+}}{l^2}
\left( 1- \frac{l^2J^2}{4\tilde{r}^4_{+} }\right)-\frac{M\sqrt{\theta}}{\tilde{r}^2_+}.
\end{eqnarray}
Again, by comparing $ \Gamma $ with the Boltzmann factor $\exp({-{E}/{{\cal T}_{H}}})$, we obtain the Hawking temperature of the noncommutative rotating BTZ black hole  
\begin{eqnarray}
{\cal T}_{H} &=&\frac{\bar{\kappa}}{4\pi}= \frac{{\cal F}^{\prime}(\tilde{r}_{+})}{4\pi},
\\
&=&\frac{2\tilde{r}_{+}}{4\pi l^2}\left( 1- \frac{l^2J^2}{4\tilde{r}^4_{+} }\right)
-\frac{M\sqrt{\theta}}{4\pi\tilde{r}^2_+}.
\end{eqnarray}
For $ \theta = 0 $ we recover the result for the Hawking temperature of the rotating BTZ black hole which is given by
\begin{eqnarray}
\label{ThJ}
	{\cal T}_{h} &=& \frac{{r}_{+}}{2\pi l^2}\left( 1- \frac{l^2J^2}{4{r}^4_{+} }\right).
\end{eqnarray}	
From Eq. (\ref{Fr1}) we obtain the mass of the noncommutative black hole, up to first order in $\sqrt{\theta}$, that is given by
\begin{eqnarray}
M&=&\frac{\tilde{r}_+^{2}}{l^{2}} +\frac{J^2}{4\tilde{r}_+^2} 
+\frac{\tilde{r}_+\sqrt{\theta}}{l^2} + \frac{\sqrt{\theta}J^2}{4 \tilde{r}^3_+} + \cdots.
\end{eqnarray} 
In order to analyze the entropy we consider the following equation:
\begin{eqnarray}
\label{ds2}
S=\int\frac{1}{{\cal T}_H}\frac{\partial M}{\partial\tilde{r}_+}d\tilde{r}_{+},
\end{eqnarray}
where \begin{eqnarray}
\label{dm2}
\frac{\partial M}{\partial\tilde{r}_+}=\frac{2\tilde{r}_+}{l^{2}}\left(1 -\frac{l^2J^2}{4\tilde{r}_+^4}\right)
+\frac{\sqrt{\theta}}{l^2}\left(1 - \frac{3l^2J^2}{4 \tilde{r}^4_+}\right)+ \cdots  .
\end{eqnarray}
The next step is to perform an expansion in $ {\cal T}_H^{-1} $ up to first order in $ \sqrt{\theta}$, 
so we can find
\begin{eqnarray}
\label{expanT}
{\cal T}_H^{-1} =4\pi\left[\frac{2\tilde{r}_{+}}{l^2}\left( 1- \frac{l^2J^2}{4\tilde{r}^4_{+} }\right)\right]^{-1}
\left\{1 +\frac{r^2_h\sqrt{\theta}}{2\tilde{r}_{+}^3}\left( 1- \frac{l^2J^2}{4\tilde{r}^4_{+} }\right)^{-1}   \right\} + \cdots.
\end{eqnarray}
Now, by replacing (\ref{dm2}) and (\ref{expanT}) in (\ref{ds2}), we obtain
\begin{eqnarray}
\hat{S}&=&4\pi\int \left\{1 +\left[\frac{r^2_h\sqrt{\theta}}{2\tilde{r}_{+}^3} 
+\frac{\sqrt{\theta}}{2\tilde{r}_{+}}\left(1 - \frac{3l^2J^2}{4 \tilde{r}^4_+}\right)
\right]\left( 1+\frac{l^2J^2}{4\tilde{r}^4_{+} }\right) + \cdots\right\}d\tilde{r}_+,
\\
&=& 4\pi\tilde{r}_{+}+2\pi\sqrt{\theta}\ln(\tilde{r}_+) -\frac{\pi r^2_h\sqrt{\theta}}{\tilde{r}_{+}^2}
-\frac{\pi r^2_h l^2J^2\sqrt{\theta}}{12\tilde{r}_{+}^6}
+\frac{2\pi l^2J^2\sqrt{\theta}}{8\tilde{r}_{+}^4}+\frac{3\pi l^4J^4\sqrt{\theta}}{64\tilde{r}_{+}^8}
+S_0 +\cdots ,
\end{eqnarray}
{where $ S_0 $  is an integration constant, and by rewriting in terms of $ r_+ $, we have }
\begin{eqnarray}
\label{entj}
\hat{S}=
4\pi{r}_{+}+ \frac{2\pi r^2_h \sqrt{\theta}}{(r^2_{-} - r^2_{+})} +2\pi\sqrt{\theta}\ln({r}_+) 
-\frac{\pi r^2_h\sqrt{\theta}}{{r}_{+}^2}
-\frac{\pi r^2_h l^2J^2\sqrt{\theta}}{12{r}_{+}^6}
+\frac{2\pi l^2J^2\sqrt{\theta}}{8{r}_{+}^4}+\frac{3\pi l^4J^4\sqrt{\theta}}{64{r}_{+}^8} + S_0
+\cdots. 
\end{eqnarray}
For $ \theta=0 $ in (\ref{entj}) we have $S=4\pi r_{+} $,  which is the entropy of the commutative rotating BTZ black hole. 
On the other hand, for  the case $ J=0 $, we have $ r_+=r_h $ and the entropy becomes 
\begin{eqnarray}\label{ThJ2}
\hat{S}&=& 4\pi\left({r}_{h}-\frac{3\sqrt{\theta}}{2} \right)+ {3\pi\sqrt{\theta}} +2\pi\sqrt{\theta}\ln({r}_h) +S_0 +\cdots.
\end{eqnarray}
Note that we have obtained a logarithmic correction for the noncommutative BTZ black hole.
{Besides, our metric corresponds to that of Ref.~\cite{Chan:1996rd} with the equivalence $ \sqrt{\theta}\equiv B $ and which is given for the non-rotating 
case ($ J=0 $)  by
\begin{eqnarray}
f_B(r) = -M +\frac{M B}{r} + \frac{r^{2}}{l^{2}},
\label{fB}
\end{eqnarray}
where $ B $ is a finite constant parameter introduced by a dilaton/scalar field. 
Hence, the horizon radius can be computed from equation (\ref{cr}) as above by taking the approximation  $B/\sqrt{l^2M}=B\sqrt{\Lambda/M} \ll1 $. So, we find
\begin{eqnarray}
r_{hb}=r_h - \frac{B}{2}+\cdots .
\end{eqnarray}
Thus, from Eq. (\ref{ds2}) a logarithmic correction is obtained for entropy, given by 
\begin{eqnarray}
{S}_B&=& 4\pi r_{hb} + 2\pi B\ln({r}_{hb})+\cdots,
\end{eqnarray}
with $\Delta S\equiv S_B-S=2\pi B\ln{r_h}+...$ associated with small (thermal) fluctuations. This approach could also be considered  
in~\cite{AdS,BTZ,Chan:1995wj,Martinez:1996gn}.
}

\section{quantum correction to the entropy}\label{HTEGUP}	
In this section in order to derive quantum corrections to the Hawking temperature and entropy of the noncommutative BTZ black hole, we will apply tunneling formalism using the Hamilton-Jacobi method. So, we will adopt the following GUP~\cite{ADV, Tawfik:2014zca, Dutta:2015, KMM, bastos}
\begin{eqnarray}
\label{gup}
\Delta x\Delta p\geq \frac{\hbar}{2}\left( 1-\frac{\alpha l_p}{\hbar} \Delta p +\frac{\alpha^2 l^2_p}{\hbar^2} (\Delta p)^2 \right),
\end{eqnarray}
where $\alpha$ is a dimensionless positive parameter and $ l_p $ is the Planck length.

In sequence, without loss of generality, we will adopt the natural units $ G=c=k_B=\hbar=l_p=1 $ and by
assuming that $ \Delta p\sim E $ 
and following the steps performed in~\cite{Anacleto:2019rfn} we can obtain the following relation for the corrected energy of the black hole 
 \begin{eqnarray}
E_{gup}\geq E\left[1-\frac{\alpha}{2(\Delta x)}+ \frac{\alpha^2}{2(\Delta x)^2}+\cdots    \right].
\end{eqnarray}
Thus, performing the same procedure as previously described, we have the following result for the probability of tunneling with corrected energy $ E_ {gup} $ given by
\begin{eqnarray}
\Gamma\simeq \exp[-2{\rm Im} ({\cal I})]=\exp\left[\frac{-4{\pi}E_{gup}}{a}\right],
\end{eqnarray}
{where $ a $ is the surface gravity}. 
Again, we compare with the Boltzmann factor and we obtain the corrected Hawking temperature of the noncommutative BTZ black hole 
\begin{eqnarray}
T\leq\tilde{T}_H\left[ 1-\frac{\alpha}{2(\Delta x)}+ \frac{\alpha^2}{2(\Delta x)^2}+\cdots   \right]^{-1}.
\end{eqnarray}
Here for simplicity we will consider the case $ J=0 $. The temperature $\tilde{T}_H$ is given by equation~(\ref{ThB}).
Furthermore, since near the event horizon of the BTZ black hole the minimum uncertainty in our model is of the order of the radius of the horizon, so the corrected temperature due to the GUP is given by
\begin{eqnarray}
\label{Tgup}
T_{gup}&\leq&\tilde{T}_H\left(1 - \frac{\alpha}{4\tilde{r}_{h}} + \frac{\alpha ^{2}}{8\tilde{r}_{h}^{2}}+\cdots\right)^{-1},
\\
&=&\frac{2\tilde{r}_h}{4\pi l^2}\left(1-\frac{r^2_h\sqrt{\theta}}{2\tilde{r}^3_h}\right)
\left(1 + \frac{\alpha}{4\tilde{r}_{h}} - \frac{\alpha ^{2}}{8\tilde{r}_{h}^{2}}+\cdots\right).
\label{tg1}
\end{eqnarray}
We can also write the result above in terms of $ r_h=l\sqrt{M} $ as follows
\begin{eqnarray}
\label{tg2}
T_{gup}\leq\frac{2{r}_h}{4\pi l^2}\left(1-\frac{\sqrt{\theta}}{2{r}_h}\right)^2
\left[1 + \frac{\alpha}{4{r}_{h}}\left(1+\frac{\sqrt{\theta}}{2r_h} +\cdots \right) 
- \frac{\alpha ^{2}}{8{r}_{h}^{2}}
\left(1+\frac{\sqrt{\theta}}{r_h}+\cdots  \right)+\cdots\right].
\end{eqnarray}
Next, we will compute the entropy of the noncommutative BTZ black hole by using the following formula:
\begin{eqnarray}
\label{sgup}
S_{gup} &=&\int \frac{1}{T_{gup}}\frac{\partial M}{\partial\tilde{r}_h}d\tilde{r}_{h},
\end{eqnarray}
where from Eq. (\ref{mrh}) we have
\begin{eqnarray}
\frac{\partial M}{\partial\tilde{r}_h}=\frac{2\tilde{r}_h}{l^{2}}\left(1 +\frac{\sqrt{\theta}}{2\tilde{r}_h}\right).
\end{eqnarray}
So, now we can obtain the corrected entropy
\begin{eqnarray}
S_{gup} &=&4\pi\int\left(1+\frac{\sqrt{\theta}}{2\tilde{r}_h} \right)
\left(1+\frac{r^2_h\sqrt{\theta}}{2\tilde{r}^3_h}\right)
\left[1 - \frac{\alpha}{4\tilde{r}_{h}} + \frac{\alpha ^{2}}{8\tilde{r}_{h}^{2}}+\cdots\right]d\tilde{r}_{h},
\\
&=&4\pi\tilde{r}_{h} +2\pi\sqrt{\theta}\ln(\tilde{r}_h)-\frac{\pi r^2_h\sqrt{\theta}}{\tilde{r}^2_h}
\nonumber\\
&-&\pi\alpha\ln(\tilde{r}_h)-\frac{\pi\alpha^2}{2\tilde{r}_h}+ \frac{\pi\alpha\sqrt{\theta}}{2\tilde{r}_h}
-\frac{\pi\alpha^2\sqrt{\theta}}{8\tilde{r}^2_h}+\frac{\pi r^2_h\alpha\sqrt{\theta}}{6\tilde{r}^3_h}
-\frac{\pi r^2_h\alpha^2\sqrt{\theta}}{16\tilde{r}^4_h} +S_0 +\cdots,
\label{sg1}
\end{eqnarray}
or by expressing the result above in terms of the $ r_h $ we have
\begin{eqnarray}
S_{gup}&=&4\pi\left( r_{h}-\frac{3\sqrt{\theta}}{2}\right) +2\pi\sqrt{\theta}\ln(r_h)+{3\pi\sqrt{\theta}}
-\pi\alpha\ln(r_h)-\frac{\pi\alpha^2}{2{r}_h}+ \frac{2\pi\alpha\sqrt{\theta}}{3{r}_h}
-\frac{3\pi\alpha^2\sqrt{\theta}}{16{r}^2_h} + S_0 +\cdots.
\end{eqnarray}
Therefore, by analyzing the result we have obtained corrections to the entropy due to the effects of GUP and also noncommutative correction.  Note that due to the effect of noncommutativity and GUP we have found logarithmic corrections for the entropy of the BTZ black hole.
For $\alpha =0 $, we have precisely the noncommutative correction to the entropy given by (\ref{ThJ2}).
{In~\cite{Jamil:2018crl}, the authors analyzed the thermodynamics of the charged rotating BTZ black hole and logarithmic corrections were also obtained for entropy in the presence of the GUP and thermal fluctuations (for small variations in $ \beta=1/T $). The logarithmic corrections to
entropy become important for very small black holes 
and negligible for very large black holes. Further studies addressing these issues were also considered in Refs.~\cite{Sadeghi:2014zna,Sadeghi:2016dvc,Pourhassan:2018wjg,Upadhyay:2017vgk}. In our case, logarithmic corrections are due to the presence of GUP and/or noncommutativity of spacetime, that  mimic small thermal fluctuations by properly identifying the corresponding  parameter to values normally found in thermal fluctuations as well discussed in Refs~\cite{Jamil:2018crl,Sadeghi:2014zna,Sadeghi:2016dvc,Pourhassan:2018wjg,Upadhyay:2017vgk}. 
}

\begin{figure}[htbh]
\centering
 \includegraphics[scale=1.0]{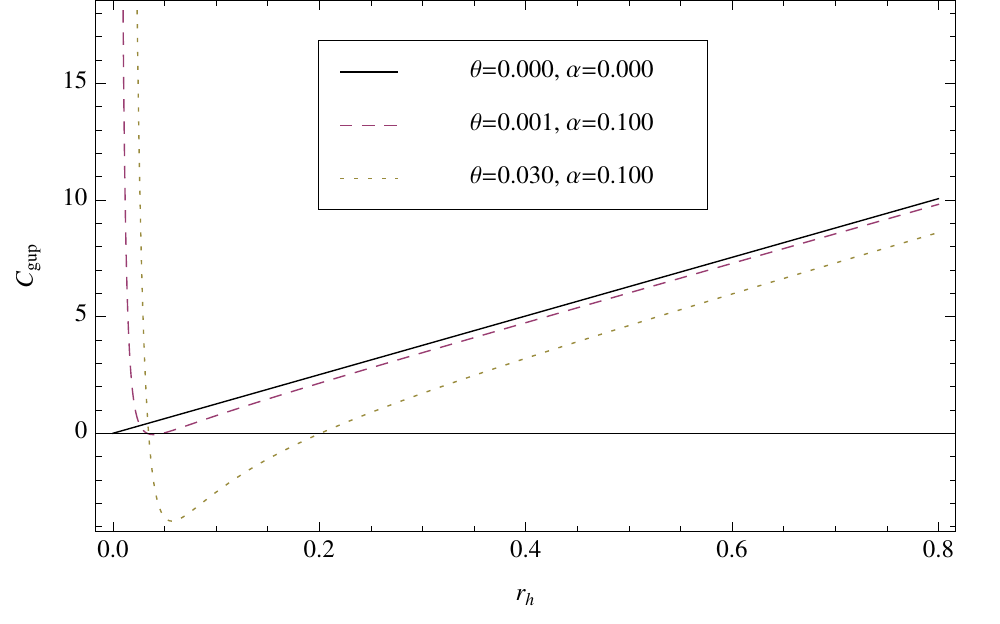}
 \caption{Specific heat capacity (Eq. (\ref{cgupb})). For $ \theta=\alpha=0 $, we have the result for the specific heat of the commutative BTZ black hole. We also show the result when 
 $ \theta\neq 0 $  and $ \alpha\neq 0 $.}
 \label{hc1}
\end{figure}
\begin{figure}[htbh]
\centering
\subfigure[]{\includegraphics[scale=0.8]{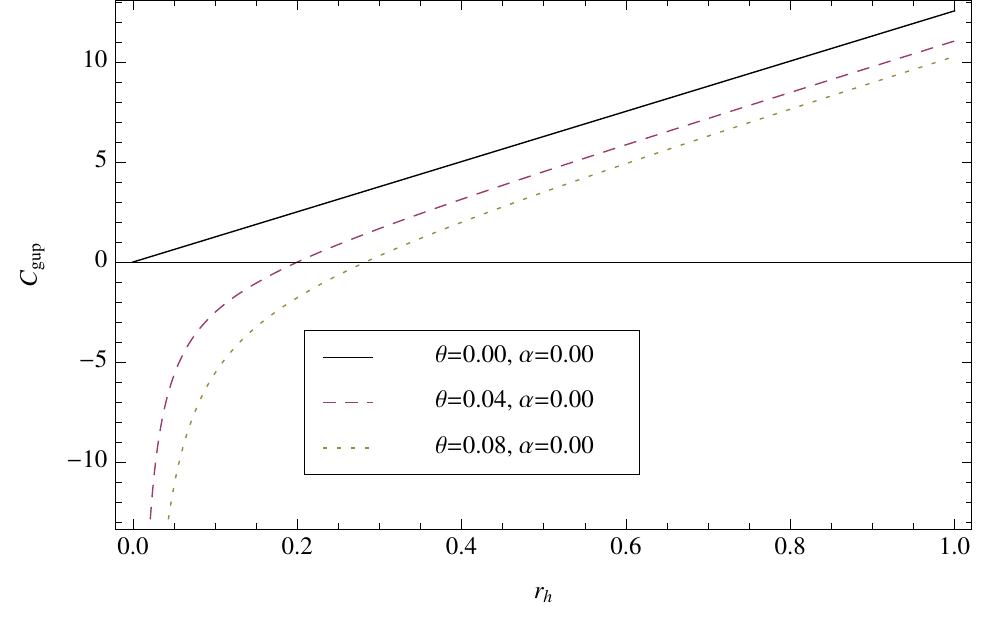}\label{hc2a}}
\qquad
\subfigure[]{\includegraphics[scale=0.8]{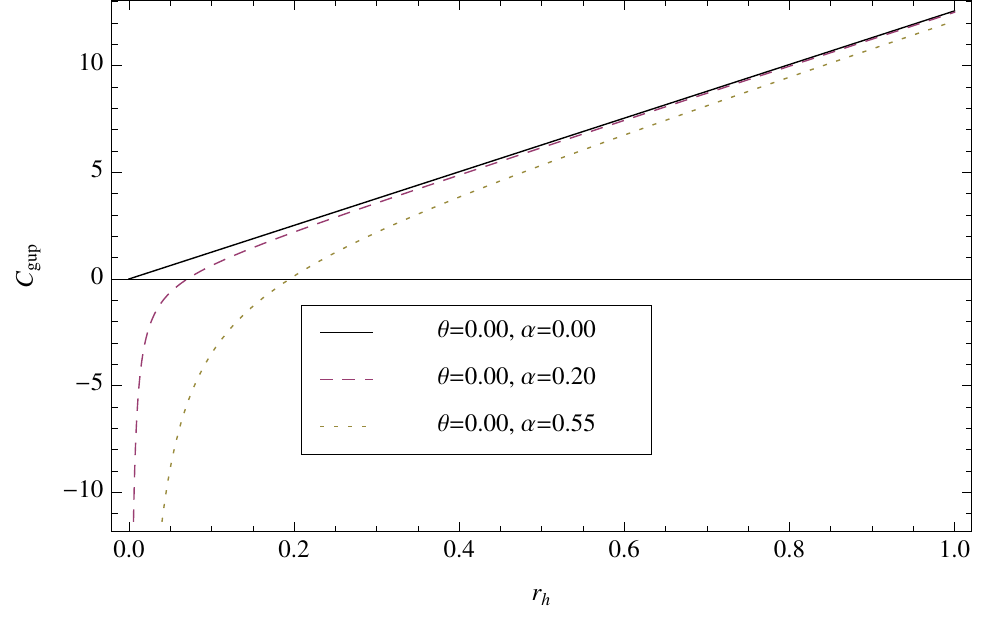}\label{hc2b}}
 \caption{Specific heat capacity. (a) For $ \theta\neq 0 $  and $ \alpha= 0 $, we have the result for the specific heat of the noncommutative BTZ black hole. (b) For $ \theta= 0 $  and $ \alpha\neq 0 $, we have the result for the specific heat of the commutative BTZ black hole with GUP.}
 \label{hc3}
\end{figure}

{At this point, we will compute Helmholtz free energy, which can be determined by using the following relationship: 
\begin{eqnarray}
F_{gup}=-\int S_{gup}\, dT_{gup}.
\end{eqnarray}
So from equations (\ref{tg1}) and (\ref{sg1}), we get
\begin{eqnarray}
F_{gup}&=&- \frac{1}{2\pi l^2}\int \left(1+ \frac{r_h^2\sqrt{\theta}}{\tilde{r}^3_h} 
+\frac{\alpha^2}{8\tilde{r}^2_h} +\frac{3r_h^2\alpha\sqrt{\theta}}{8\tilde{r}^4_h} 
-\frac{r_h^2\alpha^2\sqrt{\theta}}{4\tilde{r}^5_h}\right)S_{gup}\, d\tilde{r}_{h},
\\
&=&-\frac{\tilde{r}^2_h}{l^2}+\frac{3r_h^2\sqrt{\theta}}{2l^2\tilde{r}_h}
+\frac{5r_h^2\alpha\sqrt{\theta}}{12l^2\tilde{r}^2_h}-\frac{29r_h^2\alpha^2\sqrt{\theta}}{96l^2\tilde{r}^3_h}
-\frac{\sqrt{\theta}\tilde{r}_h}{l^2}\ln(\tilde{r}_h)+\frac{\sqrt{\theta}\tilde{r}_h}{l^2}
\nonumber\\
&+&\frac{\alpha^2\sqrt{\theta}\ln(\tilde{r}_h)}{8l^2\tilde{r}_h}+\frac{\alpha^2\sqrt{\theta}}{16l^2\tilde{r}_h}
+\frac{\alpha\tilde{r}_h}{2l^2}\ln(\tilde{r}_h)-\frac{\alpha\tilde{r}_h}{2l^2}
-\frac{r^2_h\alpha^2\sqrt{\theta}\ln(\tilde{r}_h)}{16l^2\tilde{r}^3_h}
-\frac{\alpha\sqrt{\theta}\ln(\tilde{r}_h)}{4l^2}+F_0 + \cdots,
\end{eqnarray}
where $ F_0 $ is an integration constant.

For $ \alpha = 0 $ (in the absence of the GUP) the Helmholtz free energy becomes
\begin{eqnarray}
F_{\theta}
&=&-\frac{\tilde{r}_h}{l^2}\left(\tilde{r}_h -\sqrt{\theta} \right) +\frac{3r_h^2\sqrt{\theta}}{2l^2\tilde{r}_h}
-\frac{\sqrt{\theta}\tilde{r}_h}{l^2}\ln(\tilde{r}_h)
+F_0 +\cdots,
\end{eqnarray}
or rewriting in terms of $ r_h $, we have
\begin{eqnarray}
\label{enh}
F_{\theta}
&=&-\frac{1}{l^2}\left(r_h - \frac{\sqrt{\theta}}{2}\right)\left(r_h - \frac{3\sqrt{\theta}}{2}\right) +\frac{3r_h\sqrt{\theta}}{2l^2}
-\frac{\sqrt{\theta}{r}_h}{l^2}\ln({r}_h) + F_0+\cdots.
\end{eqnarray}
}
{The correction of the specific heat capacity is given by
\begin{eqnarray}
\label{cgup}
{C}_{gup}&=&\frac{\partial M}{\partial {T}_{gup}}
=\frac{\partial M}{\partial\tilde{r}_h}\left(\frac{\partial {T}_{gup}}{\partial\tilde{r}_h}  \right)^{-1},
\\
&=&4\pi\tilde{r}_h\left(1 +\frac{\sqrt{\theta}}{2\tilde{r}_h}\right)
\left(1- \frac{\sqrt{\theta}}{\tilde{r}_h} 
-\frac{\alpha^2}{8\tilde{r}^2_h} -\frac{3\alpha\sqrt{\theta}}{8\tilde{r}^2_h} 
+\frac{\alpha^2\sqrt{\theta}}{4\tilde{r}^3_h}+\cdots\right),
\label{cgupb}
\end{eqnarray}
Now if $ \alpha=0 $ in Eq. (\ref{cgupb})  we have 
\begin{eqnarray}
\label{ct}
{C}_{\theta}=4\pi\tilde{r}_h\left(1 +\frac{\sqrt{\theta}}{2\tilde{r}_h}\right)
\left(1- \frac{\sqrt{\theta}}{\tilde{r}_h}\right)+ \cdots,
\end{eqnarray}
which in terms of $ r_h $, becomes
\begin{eqnarray}
\label{ctb}
{C}_{\theta}=4\pi{r}_h\left(1 +\frac{\sqrt{\theta}}{2{r}_h}\right)
\left(1- \frac{3\sqrt{\theta}}{2{r}_h}\right)+\cdots,
\end{eqnarray}
For $ \theta=0 $ we have, $C=4\pi{r}_h$, which is the specific heat for the commutative BTZ black hole.
Note that the specific heat vanishes at the point $r_{h} ={3\sqrt{\theta}}/2$ 
(or $ \tilde{r}_h=\sqrt{\theta}$ in Eq.~\eqref{ct} ).
In this case, we have a minimum radius 
\begin{eqnarray}
r_{\theta min}=\sqrt{l^2M_{\theta min}}=\frac{3\sqrt{\theta}}{2},
\end{eqnarray}
and then the noncommutative black hole reaches a minimum mass given by
\begin{eqnarray}
M_{\theta min}=\frac{9\theta}{ 4l^2}. 
\end{eqnarray}
Thus, this result indicates that the black hole ceases to evaporate completely and becomes a remnant. 
Next, we obtain the temperature of the remnant of the black hole by replacing the 
$ r_h \rightarrow r_{\theta min} $ 
in (\ref{tht}) 
\begin{eqnarray}
\label{thm}
{T}_{\theta rem}\approx\frac{\tilde{T}_H}{3}=\frac{r_{\theta min}}{6\pi l^2}=\frac{\sqrt{\theta}}{4\pi{l}^2}.
\end{eqnarray}	
Furthermore, from equations \eqref{ThJ2} and \eqref{enh} for $ r_h \rightarrow r_{\theta min}=3\sqrt{\theta}/2 $ we find
\begin{eqnarray}
{S}_{rem}\approx 3\pi\sqrt{\theta} + 2\pi\sqrt{\theta}\ln(3\sqrt{\theta}/2) +S_0 \approx 0, 
\qquad  S_0=-3\pi\sqrt{\theta} - 2\pi\sqrt{\theta}\ln(3\sqrt{\theta}/2),
\end{eqnarray}
and
\begin{eqnarray}
F_{rem}\approx 0 + {\cal O}(\theta) + F_0, \qquad F_0=0.
\end{eqnarray}
Hence, we have that entropy and Helmholtz free energy are zero for the remnant of the noncommutative BTZ black hole.

Now to analyze the effect of the GUP, we consider the case where $ \theta = 0 $ and $ \alpha\neq 0 $. Thus, from equation (\ref{cgupb}) we have the following contribution to specific heat:
\begin{eqnarray}
{C}_{\alpha}=4\pi r_h
\left(1- \frac{\alpha^2}{8{r}^2_h}\right)+ \cdots,
\end{eqnarray}
and the specific heat vanishes at the point $r_{h}=r_{\alpha min} ={\alpha}/2\sqrt{2}$. 
Hence, the BTZ black hole with GUP becomes a remnant with a minimum mass, 
$ M_{\alpha min}=\alpha^2/8l^2 $,  and a temperature given by
\begin{eqnarray}
\label{tarm}
T_{\alpha rem}=\frac{(T_{gup})\mid_{\theta=0}}{2}=\frac{\alpha}{8\pi l^2}.
\end{eqnarray}
Moreover, entropy and Helmholtz free energy are zero for the remnant of the BTZ black hole with GUP with
\begin{eqnarray}
{S}_{\alpha rem}\approx -\pi\alpha\ln(\sqrt{2}\alpha/4) +S_0 \approx 0, 
\qquad  S_0=\pi\alpha\ln(\sqrt{2}\alpha/4),
\end{eqnarray}
and
\begin{eqnarray}
F_{\alpha rem}\approx -(1+\sqrt{2})\frac{\alpha^2}{8l^2}- \sqrt{2}\alpha^2\ln(\sqrt{2}\alpha/4)+ F_0 \approx 0, 
\qquad F_0=(1+\sqrt{2})\frac{\alpha^2}{8l^2}+ \sqrt{2}\alpha^2\ln(\sqrt{2}\alpha/4).
\end{eqnarray}
For $ \theta\neq 0 $ and $ \alpha\neq 0 $ we can write equation (\ref{cgupb}) as follows:
\begin{eqnarray}
C_{gup}=4\pi\tilde{r}_h\left(1 +\frac{\sqrt{\theta}}{2\tilde{r}_h}\right)
\left(1 - \frac{r_{m+}}{\tilde{r}_h}\right) \left(1 - \frac{r_{m-}}{\tilde{r}_h}\right)+\cdots,
\end{eqnarray}
where
\begin{eqnarray}
r_{m\pm}=\frac{\sqrt{\theta}}{2}\pm \frac{1}{2}\sqrt{\theta + \frac{\alpha(\alpha +3\sqrt{\theta})}{2}}.
\end{eqnarray}
Note that for $\tilde{r}_h= r_{m+} $ (or $ r_h=\frac{\sqrt{\theta}}{2}+r_{m+} $) the specific heat vanishes.
The results obtained previously are recovered when $ \theta\neq 0 $ and $ \alpha=0 $ 
(or $ \theta= 0 $ and $ \alpha\neq 0 $).
For the condition of forming a remnant of a noncommutative BTZ black hole, we can write the following approximations for the minimum radius:
\begin{eqnarray}
r_{min}=\frac{\sqrt{\theta}}{2}+r_{m+}=
\begin{cases}
r_{1min}=\dfrac{3\sqrt{\theta}}{2}+\dfrac{3\alpha}{8}+{\cal O}(\alpha^2),
\\
\\
r_{2min}=\dfrac{\alpha}{2\sqrt{2}}+\dfrac{(3\sqrt{2}+8)}{8}\sqrt{\theta}+{\cal O}(\theta).
\end{cases}
\end{eqnarray}
By applying the minimum radius $ r_{min} $  
the specific heat, entropy and Helmholtz free energy are null and by (\ref{tg2}) the temperature is given by
\begin{eqnarray}
T_{\alpha\theta rem}&=&(T_{gup})\mid_{r_h=r_{min}},
\\
&=&\left(\frac{{r}_{min}}{2\pi l^2}-\frac{\sqrt{\theta}}{2\pi l^2}\right)
\left[1 + \frac{\alpha}{4{r}_{min}}+\frac{\alpha}{8{r}^2_{min}}\left(\sqrt{\theta}-\alpha \right) - \frac{\alpha ^{2}\sqrt{\theta}}{8{r}_{min}^{3}}+\cdots\right].
\end{eqnarray}
We can obtain approximate expressions for the temperature of the remnant of the noncommutative BTZ black hole by expanding it in $ \alpha $ and $ \theta $. 
So, by applying the minimum radii $ r_{1min} $ and $ r_{2min} $, the temperatures are given respectively by
\begin{eqnarray}
T_{1rem}=(T_{gup})\mid_{r_{min}=r_{1min}}=\frac{\sqrt{\theta}}{4\pi l^2}+\frac{35\alpha}{144\pi l^2}
+{\cal O}(\alpha^2),
\end{eqnarray}
and 
\begin{eqnarray}
T_{2rem}=(T_{gup})\mid_{r_{min}=r_{2min}}=\frac{\alpha}{8\pi l^2}+\frac{\sqrt{2\theta}}{4\pi l^2}
+\frac{\sqrt{\theta}}{\pi l^2}+{\cal O}(\theta).
\end{eqnarray}
Then, by doing $ \alpha=0 $ in $ T_{1rem} $ we obtain the result of (\ref{thm}), and for $ \theta=0 $ 
in $ T_{2rem} $ we recover the result of (\ref{tarm}).

Now in order to verify the stability of the black hole, we show in Figs. \ref{hc1}  and \ref{hc3} the specific heat behavior.
In Fig. \ref{hc1} we show that the specific heat is positive for $ \theta =0.001 $ and $ \alpha=0.1 $, and so the noncommutative BTZ black hole with GUP is stable.
In addition, we observed that specific heat vanishes to a critical radius.
Furthermore, for $ \theta =0.03 $ and $ \alpha=0.1 $ one achieves two points where the
specific heat vanishes, with an unphysical region in between. 

In Fig~\ref{hc2a} we verify the behavior of the specific heat for $ \theta \neq 0 $ and $ \alpha =0 $, 
and in Fig.~\ref{hc2b} for $ \theta = 0 $ and $ \alpha\neq 0 $. 
Note that the specific heat vanishes before entering into an unphysical zone. 
The BTZ black hole decreases its size until achieve a critical radius where it ceases to evaporate and becomes a remnant of the noncommutative BTZ black hole.
}

\section{CONCLUSIONS}	\label{concl}
In summary, we have considered the metric of a noncommutative BTZ black hole implemented via Lorentzian mass distribution. Thus, applying the Hamilton-Jacobi approach and the WKB approximation we have obtained noncommutative corrections to Hawking temperature and entropy. 
In addition, we have found a logarithmic correction to the entropy of the  BTZ black hole due to the effect of noncommutativity. 
We also have verified the stability of the BTZ black hole by calculating the specific heat capacity 
and have shown that the noncommutative BTZ black hole becomes a remnant with a minimum mass 
$ M_{\theta min}={9\theta}/{4 l^2} $. 
Therefore, the contribution of the noncommutative corrections introduces a GUP effect.
We also investigated the effect of GUP by calculating Hawking temperature and entropy of the noncommutative BTZ black hole. Due to the effect of noncommutativity and GUP we have found a logarithmic corrections for the entropy of the BTZ black hole, in the form $S_{gup}\sim S+(c_1+c_2)\ln{S}+...$, where the `species' $c_i=(-\alpha, \theta)$ are essentially related to each corresponding parameter of correction.

\acknowledgments
We would like to thank CNPq, CAPES and CNPq/PRONEX/FAPESQ-PB (Grant nos. 165/2018 and 015/2019),  for partial financial support. MAA, FAB and EP acknowledge support from CNPq (Grant nos. 306962/2018-7 and  433980/2018-4, 312104/2018-9, 304852/2017-1).

\end{document}